\documentclass[journal]{IEEEtran}
\usepackage{graphicx,amssymb,amstext,amsmath}
\usepackage{amsthm}
\usepackage{color}
\usepackage[none]{hyphenat}

\theoremstyle{definition}

\usepackage{cite}
\usepackage{amsmath}
\usepackage{algorithm}
\usepackage{algorithmic}
\usepackage{subfigure}
\usepackage{multirow}
\usepackage{makecell}
\usepackage{epstopdf}

\usepackage{soul}
\soulregister\cite7

\begin{document}

\title{\LARGE An Improved EPA based Receiver Design for Uplink LDPC Coded SCMA System}

\author{Lingyun~Chai, \IEEEmembership{Graduate Student Member,~IEEE,}
        Zilong~Liu, \IEEEmembership{Senior Member,~IEEE,}
        Pei~Xiao, \IEEEmembership{Senior Member,~IEEE,}
        Amine~Maaref, \IEEEmembership{Senior Member,~IEEE,}
        and Lin~Bai, \IEEEmembership{Senior Member,~IEEE}
        \thanks{
        Lingyun Chai is with the School of Electronic and Information Engineering, Beihang University, Beijing 100191, China (e-mail: lingyunchai@buaa.edu.cn).

        Zilong Liu is with the School of Computer Science and Electronics Engineering, University of Essex, UK (e-mail: zilong.liu@essex.ac.uk).

        Pei Xiao is with 5GIC  \& 6GIC, Institute for Communication Systems (ICS), University of Surrey, UK (e-mail: p.xiao@surrey.ac.uk).

        Amine Maaref is with the Huawei Technologies Canada Co. Ltd., Canada (e-mail: amine.maaref@huawei.com).

        Lin Bai is with the School of Cyber Science and Technology, Beihang University, Beijing 100191, China, and also with the Beijing Laboratory for General Aviation Technology, Beihang University, Beijing 100191, China (e-mail: l.bai@buaa.edu.cn).}
}

\maketitle

\begin{abstract}
Sparse code multiple access (SCMA) is an emerging paradigm for efficient enabling of massive connectivity in future machine-type communications (MTC). In this letter,  we conceive the uplink transmissions of the low-density parity check (LDPC) coded SCMA system. Traditional receiver design of LDPC-SCMA system, which is based on message passing algorithm (MPA) for multiuser detection followed by individual LDPC decoding, may suffer from the drawback of the high complexity and large decoding latency, especially when the system has large codebook size and/or high overloading factor. To address this problem, we introduce a novel receiver design by applying the expectation propagation algorithm (EPA) to the joint detection and decoding (JDD) involving an aggregated factor graph of LDPC code and sparse codebooks. Our numerical results demonstrate the superiority of the proposed EPA based JDD receiver over the conventional Turbo receiver in terms of both significantly lower complexity and faster convergence rate without noticeable error rate performance degradation.
\end{abstract}

\IEEEpeerreviewmaketitle

\section{Introduction}

SCMA is a special class of code-domain non-orthogonal multiple access (NOMA) \cite{Liu2021}, in which multiple users are separated by assigning distinctive sparse codebooks \cite{Nikopour2013Sparse}. Unlike its predecessor -- low-density signature (LDS) which relies on sparse sequence spreading of constellation symbols \cite{Hoshyar2008Novel,Liu2019}, several bits of every SCMA user are directly mapped into a multi-dimensional complex codeword selected from a carefully designed sparse codebook \cite{Taherzadeh2014}. By taking advantage of the sparsity of SCMA, MPA has been used for near optimal error rate performance as well as lower complexity than the maximum likelihood detection.

Inspired by the Turbo receiver principle, an iterative multiuser receiver has been developed for a coded SCMA system to achieve improved error rate performance by fully exploiting the system diversity gain and coding gain \cite{Wu2015Iterative}. Besides, significant efforts have been made aiming to lower the receiver complexity. As an example, \cite{Wei2017Low} proposed a low complexity decoding algorithm in which list sphere decoding (LSD) is employed for efficient pruning of the search branches in the corresponding tree.
For an SCMA system with multiple receive antennas, \cite{Meng2017Low} studied a low complexity receiver based on EPA whose complexity grows linearly with $M$ and $d_f$, where $M$ and $d_f$ denote the codebook size and the number of active users superimposed over each resource node, respectively. This is in sharp contrast to the decoding complexity of the MPA detector which is in the order of $M^{d_f}$, i.e., $O(M^{d_f})$.

There have been some works on merging the SCMA detection and LDPC decoding under a joint sparse graph (JSG). Their main idea is to pass the belief messages over an aggregated factor graph associated to both the sparse codebooks/sequences and LDPC codes \cite{Wen2015Design}. It is shown that their design leads to a low-complexity JDD with excellent performance gain compared to the traditional Turbo structured LDS-OFDM receiver. Recently, similar ideas have been reported in \cite{Han2016High,Lai2018Subgraph,Pan2018Design} by the joint trellis-based receiver of LDPC and polar coded SCMA systems.

Motivated by the research works in \cite{Meng2017Low}, \cite{Han2016High} and to further reduce the receiver complexity, we propose to apply the EPA detector to the JDD receiver for uplink LDPC coded SCMA system. Compared to \cite{Meng2017Low} and \cite{Han2016High}, our proposed receiver design brings in performance gains through combining the JDD method and replacing the MPA detector with the EPA detector, respectively.
Specifically, we present an improved EPA detector whose propagation of belief messages can be bypassed or early terminated once a certain threshold value associated with those belief messages is satisfied. Such a design can effectively reduce the whole EPA detection complexity with negligible error rate performance degradation.
On the one hand, simulation results show that the proposed JDD/EPA receiver is able to attain significant performance gains and faster convergence in reference to the traditional Turbo receiver \cite{Meng2017Low} with comparable complexity. On the other hand, the JDD/EPA receiver can effectively lower the complexity with slight error rate loss compared to the JDD/MPA receiver in \cite{Han2016High}.

\emph{Notations:} Bold uppercase, bold lowercase and lowercase letters denote matrices, vectors and scalars, respectively. ${\left(  \cdot  \right)^{\rm{T}}}$ indicates the matrix or vector transpose. ${\rm{diag}}\left( {{a_1},...,{a_N}} \right)$ denotes a diagonal matrix with diagonal elements ${{a_1},...,{a_N}}$. The notation ${\left(  \cdot  \right)_{i,j}}$ is used to indicate the $i$-th row and $j$-th column element of the matrix. $x \sim {\cal C}{\cal N}\left( {\mu ,\zeta } \right)$ indicates that the random variable $x$ follows a complex Gaussian distribution with mean $\mu$ and variance $\zeta$.

\section{System Model}
An LDPC coded SCMA system is considered with $K$ independent mobile users (each equipped with a single transmit antenna) multiplexed on $N$ orthogonal resource nodes, where the base station consists of $N_r$ receive antennas. Each orthogonal resource node is essentially a subcarrier channel in a multicarrier system. To support overloaded multiuser communication, we have $K>N$, where the overloading factor is defined as $\lambda = K/N$.

For user $k\in\{1,2,\cdots,K\}$, its SCMA encoder maps coded bits ${{\bf{c}}_k}$ to a multi-dimensional complex codeword ${{\bf{x}}_k}$ of length $N$. The codeword ${{\bf{x}}_k}$ is drawn from SCMA codebook ${\mathcal{C}_k}$ with cardinality of $M$. Therefore, each codebook may be written as an $N\times M$ matrix. For each user, $Q = {\log _2}M$ coded bits are mapped to an $N$-dimensional SCMA codeword. Let ${{\bf{c}}_k} = {\left[ {{c_{k,1}},{c_{k,2}}, \cdots ,{c_{k,Q}}} \right]^{\rm{T}}}$ be the coded bits, and ${{\bf{x}}_k} = {\left[ {{x_{k,1}},{x_{k,2}}, \cdots ,{x_{k,N}}} \right]^{\rm{T}}}$ be the corresponding SCMA codeword. These SCMA codewords are sparse with ${d_r} < N$ non-zero elements and the sparsity of codewords is exploited at the receiver through MPA or EPA detection.

Let ${\bf{H}}_k^{{n_r}} = {\rm{diag}}\left( {h_{k,1}^{{n_r}},h_{k,2}^{{n_r}}, \cdots ,h_{k,N}^{{n_r}}} \right)$ denote the channel gain matrix between the $k$-th user and the $n_r$-th receive antenna at the base station.
Then the corresponding received signal in the uplink can be expressed as
\begin{equation}\label{1}
{{\bf{y}}^{{n_r}}} = \sum\limits_{k = 1}^K {{\bf{H}}_k^{{n_r}}{{\bf{x}}_k} + {{\bf{n}}^{{n_r}}}},
\end{equation}
where ${{\bf{y}}^{{n_r}}} = {\left[ {y_1^{{n_r}},y_2^{{n_r}}, \cdots ,y_N^{{n_r}}} \right]^{\rm{T}}}$ is received signal vector at the $n_r$-th antenna, and ${{\bf{n}}^{{n_r}}}$ denotes the additive complex Gaussian noise vector whose distribution is characterized by ${\cal C}{\cal N}(0,{\sigma _n^2})$.

Let us denote by ${d_r}$ the number of resource nodes occupied by each user, and ${d_f}$ the number of active users superimposed over a resource node. Additionally, $\mathcal{F}(n) = \left\{ {k :{x_{k,n}} \ne 0} \right\}$ denotes the index set of the neighboring SCMA variable nodes (VNs) which are connected with function node (FN) $n$ for each receive antenna, $\mathcal{V}(k) = \left\{ {n:  {x_{k,n}} \ne 0} \right\}$ denotes the index set of the neighboring FNs which are connected with the VN $k$. The cardinalities of $\mathcal{V}(k)$ and $\mathcal{F}(n)$ satisfy the following individual constraints: $\left| \mathcal{V}(k) \right| = {d_r},k = 1,2,...,K$ and $\left|\mathcal{F}(n) \right| = {d_f},n = 1,2,...,N$.
It is noted that in an OFDM based SCMA system, every FN corresponds to a subcarrier channel, whereas every VN corresponds to the transmit bits of an SCMA user.

\section{Improved EPA Based JDD Receiver}

EPA is a Bayesian inference method which aims for approximating the sophisticated \textit{a posterior} belief distribution with a tractable distribution, such as Gaussian distribution through moment matching\cite{Meng2017Low}. Define the projection of a particular distribution $p$ into some distribution set $\Phi $ as
\begin{equation}\label{2}
{\rm{Pro}}{{\rm{j}}_\Phi }\left( p \right) = \arg \mathop {\min }\limits_{q \in \Phi } D\left( {p||q} \right),
\end{equation}
where $D\left( {p||q} \right)$ denotes the Kullback-Leibler (KL) divergence which is used to measure the similarity between the \textit{a posterior} belief distribution and the tractable distribution after projection.

The complexity of MPA is in the order ${\cal O}\left( {{M^{{d_f}}}} \right)$, which may be prohibitively high when $M$ and/or $d_f$ are large.
A major advantage of the EPA detector is its low complexity, which scales linearly with the values of the $M$ and ${d_f}$. This makes the EPA detector very attractive for practical implementation. In this section, an improved EPA based JDD receiver is designed by updating the belief messages iteratively on the joint sparse graph.
The joint sparse graph of the JDD receiver for an uplink LDPC coded SCMA system is depicted in Fig. 1, where belief messages are exchanged between SCMA VNs and LDPC VNs by merging the factor graphs of SCMA and LDPC into an integrated one, ${\Pi _k}$ and $\Pi _k^{ - 1}$ denote the interleaver and deinterleaver of user $k$, respectively.
In the sequel, we will discuss in detail how the belief messages are updated and transferred between different nodes in the joint sparse graph.
\begin{figure}[htbp]
  \centering
  \includegraphics[scale=0.75]{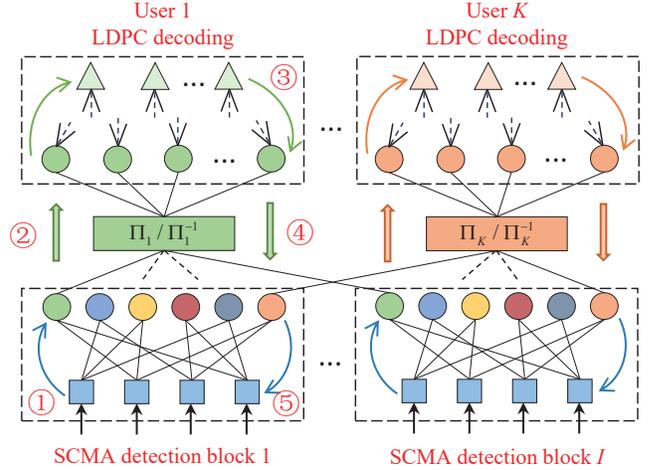}
  \caption{Illustration of the joint sparse graph of the JDD receiver for LDPC-SCMA system.}
\end{figure}

\subsubsection{SCMA Function Nodes Update}
In every SCMA data block, let $L_{F_n^{{n_r}} \to {V_k}}^t$ denote the LLR passed from FN $F_n^{{n_r}}$ to VN ${V_k}$ at the $n_r$ receive antenna in the $t$-th iteration, and $L_{{V_k} \to F_n^{{n_r}}}^t$ the log-likelihood ratio (LLR) of the opposite direction. According to the EPA principle \cite{Meng2017Low}, the mean and variance update equations of the belief message $I_{F_n^{{n_r}} \to {V_k}}^t$ are described as follows:
\begin{equation}\label{3}
\mu _{F_n^{{n_r}} \to {V_k}}^t = \frac{1}{{h_{k,n}^{{n_r}}}}\left( {y_n^{{n_r}} - \sum\limits_{k' \in \mathcal{F}(n)\backslash k} {h_{k',n}^{{n_r}}\mu _{{V_{k'}} \to F_n^{{n_r}}}^{t-1}} } \right),
\end{equation}
\begin{equation}\label{4}
\zeta _{F_n^{{n_r}} \to {V_k}}^t = \frac{1}{{{{\left| {h_{k,n}^{{n_r}}} \right|}^2}}}\left( {{\sigma _n^2} + \sum\limits_{k' \in \mathcal{F}(n)\backslash k} {{{\left| {h_{k',n}^{{n_r}}} \right|}^2}\zeta _{{V_{k'}} \to F_n^{{n_r}}}^{t-1}} } \right),
\end{equation}
\begin{equation}\label{5}
L_{F_n^{{n_r}} \to {V_k}}^t\left( {\hat m} \right) =  - \frac{{{{\left| {{x_{k,n}}\left( {\hat m} \right) - \mu _{F_n^{{n_r}} \to {V_k}}^t} \right|}^2}}}{{\zeta _{F_n^{{n_r}} \to {V_k}}^t}}.
\end{equation}

\subsubsection{Generate the extrinsic LLRs from SCMA detector}
The updated LLRs of VNs from neighboring FNs are regarded as the extrinsic information of the SCMA detector which can be used as the \textit{a priori} LLRs for the LDPC decoder. Thus,
\begin{equation}\label{6}
L_k^{e,{\rm{SCMA}}}\left( {\hat m} \right) = \sum\limits_{{n_r} = 1}^{{N_r}} {\sum\limits_{n \in \mathcal{V}(k)} {L_{F_n^{{n_r}} \to {V_k}}^t\left( {\hat m} \right)} } .
\end{equation}

For illustration, in the simulation part, we consider $M=4$. Thus, every two coded bits ${c_k} \in \{ 00,01,10,11\} $ are mapped into an $N$-dimensional SCMA codeword ${\mathbf{x}_{k}} \in \{ {\mathbf{x}_{k}}(1), \ldots ,{\mathbf{x}_{k}}(4)\} $.
It should be pointed out that for message passing from SCMA VNs to LDPC VNs, one needs to transform symbol LLRs to bit LLRs.
Let us assume that there are $I$ SCMA symbols to be sent for each user. For the $i$-th $\left(i \in \left\{ {1,...,I} \right\}\right)$ SCMA block, the symbol-to-bit LLRs transformation can be calculated by applying Max-Log approximation as follows:
\begin{equation}\label{7}
\left\{ {\begin{array}{*{20}{l}}
\begin{array}{c}
\begin{aligned}
L_{k,2i - 1}^{e,{\rm{SCMA}}}(\hat b) = & \max \left( \scriptstyle{{L_{k,i}^{e,{\rm{SCMA}}}\left( {\hat m = 1} \right),L_{k,i}^{e,{\rm{SCMA}}}\left( {\hat m = 2} \right)}} \right)\\
& - \max \left( \scriptstyle{{L_{k,i}^{e,{\rm{SCMA}}}\left( {\hat m = 3} \right),L_{k,i}^{e,{\rm{SCMA}}}\left( {\hat m = 4} \right)}} \right)
\end{aligned}
\end{array}\\
\begin{array}{c}
\begin{aligned}
L_{k,2i}^{e,{\rm{SCMA}}}(\hat b) = & \max \left( \scriptstyle{{L_{k,i}^{e,{\rm{SCMA}}}\left( {\hat m = 1} \right),L_{k,i}^{e,{\rm{SCMA}}}\left( {\hat m = 3} \right)}} \right)\\
& - \max \left( \scriptstyle{{L_{k,i}^{e,{\rm{SCMA}}}\left( {\hat m = 2} \right),L_{k,i}^{e,{\rm{SCMA}}}\left( {\hat m = 4} \right)}} \right)
 \end{aligned}
\end{array}
\end{array}} \right..
\end{equation}

Then the extrinsic bit LLRs of user $k$ are deinterleaved as the \textit{a priori} LLRs input to its corresponding LDPC decoder, i.e.,
\begin{equation}\label{8}
L_k^{a,{\rm{LDPC}}}({\bf{\hat b}}) = {\Pi ^{ - 1}} \left( {L_k^{e,{\rm{SCMA}}}({\bf{\hat b}})} \right).
\end{equation}

\subsubsection{LDPC decoder}
We assume that the LDPC parity check matrix includes $J$ LDPC VNs and $L$ check nodes (CNs). For each user, the updated LLR to an LDPC VN consists of the \textit{a priori} LLR $L_{k,j}^{a,{\rm{LDPC}}}$ and the LLRs passed from the corresponding CNs $L_{{C_{l,k}} \to L{V_{j,k}}}$,
\begin{equation}\label{9}
L_{L{V_{j,k}}}^t(\hat b) = L_{k,j}^{a,{\rm{LDPC}}}(\hat b) + \sum\limits_{l \in {\mathcal{N}_v}(j)} {L_{{C_{l,k}} \to L{V_{j,k}}}^{t - 1}(\hat b)},
\end{equation}
where ${\mathcal{N}_v}(j)$ is the set of all the CNs connected to LDPC VN ${L{V_{j}}}$.
Within each iteration loop, the update from an LDPC VN to a CN dynamically takes into account the belief messages from the other CNs and the corresponding SCMA VN which are connected to that specific LDPC VN. It should be noted that in traditional Turbo receiver, the update from an LDPC VN to a CN is carried out internally within the LDPC decoder and does not consider the relevant SCMA VN except for its extrinsic information at the beginning of LDPC decoding.
Then the conventional LDPC decoding method such as normalized min-sum algorithm is utilized to update the CNs. Furthermore, layered belief propagation (layered BP) decoding can be adopted to accelerate the convergence rate and improve the decoding performance.

\subsubsection{Generate the \textit{a priori} LLRs for SCMA detector}
The updated extrinsic bit LLRs ${L_k^{e,{\rm{LDPC}}}({\bf{\hat b}})}$ can be obtained from its corresponding connected LDPC CNs, which will be sent to the corresponding interleaver.
After the interleaving, the \textit{a priori} bit LLRs of the SCMA detector (which is EPA in this work) can be expressed as
\begin{equation}\label{10}
L_k^{a,{\rm{SCMA}}}({\bf{\hat b}}) = \Pi \left( {L_k^{e,{\rm{LDPC}}}({\bf{\hat b}})} \right).
\end{equation}

To pass the belief messages from LDPC VNs to SCMA VNs, one needs to transform bit LLRs to symbol LLRs.
For the $i$-th $\left(i \in \left\{ {1,...,I} \right\}\right)$ SCMA block, the \textit{a priori} symbol-level LLRs can be calculated as
\begin{equation}\label{11}
\begin{array}{*{20}{l}}
L_{k,i}^{a,{\rm{SCMA}}}\left( {\hat m} \right) =
{\left\{ {\begin{array}{*{20}{l}}
\begin{aligned}
\scriptstyle{ 0.5 \cdot L_{k,2i - 1}^{a,{\rm{SCMA}}}( {\hat b} ) + 0.5 \cdot L_{k,2i }^{a,{\rm{SCMA}}}( {\hat b} ),\;\;{\hat m} = 1}\\
\scriptstyle{ 0.5 \cdot L_{k,2i - 1}^{a,{\rm{SCMA}}}( {\hat b} ) - 0.5 \cdot L_{k,2i }^{a,{\rm{SCMA}}}( {\hat b} ),\;\;{\hat m} = 2}\\
\scriptstyle{-0.5 \cdot L_{k,2i - 1}^{a,{\rm{SCMA}}}( {\hat b} ) + 0.5 \cdot L_{k,2i }^{a,{\rm{SCMA}}}( {\hat b} ),\;\;{\hat m} = 3}\\
\scriptstyle{-0.5 \cdot L_{k,2i - 1}^{a,{\rm{SCMA}}}( {\hat b} ) - 0.5 \cdot L_{k,2i }^{a,{\rm{SCMA}}}( {\hat b} ),\;\;{\hat m} = 4}
\end{aligned}
\end{array}} \right.}
\end{array}.
\end{equation}

\subsubsection{SCMA Variable Nodes Update}
Similarly, the update from an SCMA VN to an FN is calculated based on belief messages from the other FNs and the corresponding LDPC VN which are connected to that specific SCMA VN.
The updated LLR to an SCMA VN consists of the \textit{a priori} LLR $L_k^{a,{\rm{SCMA}}}$ and all the relevant ${L_{F_n^{{n_r}} \to {V_k}}}$,
\begin{equation}\label{12}
\begin{array}{l}
L{q^t}\left( {{{\bf{x}}_k}(\hat m)|{\bf{y}}} \right) = {\rm{ }}L_k^{a,{\rm{SCMA}}}(\hat m) + \sum\limits_{{n_r} = 1}^{{N_r}} {\sum\limits_{n \in \mathcal{V}(k)} {L_{F_n^{{n_r}} \to {V_k}}^{t}\left( {\hat m} \right)} } ,
\end{array}
\end{equation}
where $m \in \left\{ {1,...,M} \right\}$. For each VN ${V_{k}}$, the \textit{a posterior} belief probability for all ${{{\bf{x}}_k} \in {{\cal C}_k}}$ can be calculated below:
\begin{equation}\label{13}
{q^t}({{\bf{x}}_k}|{\bf{y}}) = \frac{\exp \left( {L{q^t}({{\bf{x}}_k}|{\bf{y}})} \right)}{{1 + \exp \left( {L{q^t}({{\bf{x}}_k}|{\bf{y}})} \right)}}.
\end{equation}

$\bullet$ Threshold aided EPA

To optimize EPA detection, the threshold aided EPA is further proposed. In VN update, denote by ${q^t}({{\bf{x}}_k}|{\bf{y}})$ the \textit{a posterior} belief probability for all ${{{\bf{x}}_k} \in {{\cal C}_k}}$. $\max \left( {{q^t}({{\bf{x}}_k}|{\bf{y}})} \right)$ denotes the maximum \textit{a posterior} belief probability corresponding to the most likely SCMA codeword of user $k$. We define the following indicator function as
\begin{equation}\label{14}
{\Psi (k)} = \left\{ {\begin{array}{*{20}{c}}
{1,\ \max \left( {{q^t}({{\bf{x}}_k}|{\bf{y}})} \right) \ge \beta }\\
{0,\ \max \left( {{q^t}({{\bf{x}}_k}|{\bf{y}})} \right) < \beta }
\end{array}} \right.,
\end{equation}
where $\beta  \in \left[ {0.5,1} \right]$. If the equation $\sum\nolimits_{k = 1}^K {\Psi \left( k \right) = K}$ holds, the message updating of FNs and VNs in the current SCMA block can be bypassed. As a result, the overall complexity of the JDD/EPA receiver can be further reduced.

According to the EPA principle \cite{Meng2017Low}, the \textit{a posterior} mean $\mu _{k,n}^{t}$ and variance $\zeta _{k,n}^{t}$ are respectively computed with the approximated \textit{a posterior} belief as follows:
\begin{equation}\label{15}
\mu _{k,n}^t = \sum\limits_{{{\bf{x}}_k} \in {\mathcal{C}_k}} {{q^t}({{\bf{x}}_k}|{\bf{y}}) \cdot {x_{k,n}}},
\end{equation}
\begin{equation}\label{16}
\zeta _{k,n}^t = \sum\limits_{{{\bf{x}}_k} \in {\mathcal{C}_k}} {{q^t}({{\bf{x}}_k}|{\bf{y}}){{\left| {{x_{k,n}} - \mu _{k,n}^t} \right|}^2}}.
\end{equation}

The variance and mean of the belief message ${I_{{V_{k}} \to {F_{n}^{{n_r}}}}}$ delivered from VN ${V_{k}}$ to FN ${F_{n}^{{n_r}}}$ is calculated as follows:
\begin{equation}\label{17}
\zeta _{{V_k} \to F_n^{{n_r}}}^t = {\left( {\frac{1}{{\zeta _{k,n}^t}} - \frac{1}{{\zeta _{F_n^{{n_r}} \to {V_k}}^{t}}}} \right)^{ - 1}},
\end{equation}
\begin{equation}\label{18}
\mu _{{V_k} \to F_n^{{n_r}}}^t = \zeta _{{V_k} \to F_n^{{n_r}}}^t\left( {\frac{{\mu _{k,n}^t}}{{\zeta _{k,n}^t}} - \frac{{\mu _{F_n^{{n_r}} \to {V_k}}^{t}}}{{\zeta _{F_n^{{n_r}} \to {V_k}}^{t}}}} \right).
\end{equation}

$\bullet$ Approximate EPA

In SCMA VNs update, the calculation of variance and mean of the belief message ${I_{{V_{k}} \to {F_{n}^{{n_r}}}}}$ includes a number of complex multiplication and division operations. With the increasing iterations, taking into account that $\zeta _{k,n}^t \to 0$ and neglecting the high-order term ${1 \mathord{\left/ {\vphantom {1 {\zeta _{F_n^{{n_r}} \to {V_k}}^t}}} \right. \kern-\nulldelimiterspace} {\zeta _{F_n^{{n_r}} \to {V_k}}^t}}$ derived in \cite{Meng2015Expectation}, where ${\zeta _{k,n}^t}$ and ${\mu _{k,n}^t}$ play a pivotal role in the above calculation, \eqref{17} and \eqref{18} can thus be directly approximated as
\begin{equation}\label{19}
\zeta _{{V_k} \to F_n^{{n_r}}}^t \approx \zeta _{k,n}^t,
\end{equation}
\begin{equation}\label{20}
\mu _{{V_k} \to F_n^{{n_r}}}^t \approx \mu _{k,n}^t.
\end{equation}

The core idea of the JDD receiver is to jointly update the extrinsic information of SCMA detection and LDPC decoding within one iteration loop.
In other words, JDD can fully utilize the latest updated extrinsic information at each iteration.
The step-by-step description of the JDD receiver with proposed improved EPA is summarized in \textbf{Algorithm 1}.

\begin{algorithm}[htbp]
\caption{Proposed improved EPA based JDD receiver}
\small
\hspace*{0.02in} {\bf Initialization:}\\
\hspace*{0.02in} {$\mu _{{{V_{k} \to F_{n}^{n_r}}}}^0 = 0$, $\zeta _{{{V_{k} \to F_{n}^{n_r}}}}^0 = \mathrm{max}$, $\beta  \in \left[ {0.5,1} \right]$, ${\Psi \left( k \right) = 0}, \forall k\in\mathcal{K}$.} \\
\hspace*{0.02in} {\bf Iteration:}
\begin{algorithmic}[1]
\FOR{$t=1:N_{iter}$}

\STATE {SCMA FNs update:}
\FOR{$i=1:I$}
\IF{$\sum\nolimits_{k = 1}^K {\Psi \left( k \right) = K}$}
\STATE $L_{F_n^{{n_r}} \to {V_k}}^t = L_{F_n^{{n_r}} \to {V_k}}^{t - 1}, i=i+1$, continue;
\ELSE
\STATE Compute $\mu _{{F_{n}^{n_r}} \to {V_{k}}}^t$ and $\zeta _{{F_{n}^{n_r}} \to {V_{k}}}^t$ via \eqref{3}-\eqref{4};
\STATE Compute $L_{{F_{n}^{n_r}} \to {V_{k}}}^t$ via \eqref{5}, $i=i+1$;
\ENDIF
\ENDFOR

\STATE {Generate the extrinsic LLRs from EPA detector\eqref{7};}
\STATE {Deinterleave LLRs using\eqref{8};}
\STATE {Layered BP LDPC decoding;}
\STATE {Interleave LLRs using\eqref{10};}
\STATE {Generate the \textit{a priori} LLRs for EPA detector\eqref{11};}

\STATE {SCMA VNs update:}
\FOR{$i=1:I$}
\STATE Compute $L{q^t}({{\bf{x}}_k}|{{\bf{y}}})$ and ${q^t}({{\bf{x}}_k}|{{\bf{y}}})$ via \eqref{12}-\eqref{13};
\IF{$\sum\nolimits_{k = 1}^K {\Psi \left( k \right) = K}$}
\STATE $\zeta _{{V_{k}} \to {F_{n}^{n_r}}}^t = \zeta _{{V_{k}} \to {F_{n}^{n_r}}}^{t-1}$, $\mu _{{V_{k}} \to {F_{n}^{n_r}}}^t = \mu _{{V_{k}} \to {F_{n}^{n_r}}}^{t-1}$,
\STATE $i=i+1$, continue;
\ELSE
\STATE Compute $\mu _{k,n}^{t}$ and $\zeta _{k,n}^{t}$ via \eqref{15}-\eqref{16};
\STATE Compute $\zeta _{{V_{k}} \to {F_{n}^{n_r}}}^t$ and $\mu _{{V_{k}} \to {F_{n}^{n_r}}}^t$ via \eqref{17}-\eqref{18},
\STATE $i=i+1$;
\ENDIF
\ENDFOR

\ENDFOR
\end{algorithmic}
{\bf Output: ${{\bf{\hat b}}_k} = ( {L_k^{e,{\rm{LDPC}}}({\bf{\hat b}})} ) < 0$}.
\end{algorithm}

\section{Numerical Results}
In this section, an uplink SCMA system is considered over the Rayleigh fading channel, where $K = 6$ single-antenna users are multiplexed over $N=4$ orthogonal resources and the receiver is equipped with $N_r=2$ antennas. The codebook size of SCMA is $M=4$ and the Huawei codebook in \cite{Nikopour2013Sparse} is used.
The irregular LDPC codes is generated by the PEG algorithm with the variable and check nodes degree distribution being $\lambda \left( x \right) = 0.1x + 0.9{x^2}$ and $\rho(x)=0.225{x^4} + 0.755{x^5} + 0.02{x^6}$, where the information bits $L=$ 200, the coded bits $J=$ 400 and code rate $R=$ 1/2. Each user has $I=$ 200 SCMA symbols to transmit.
The maximum iterations of the JDD receiver are $N_{\rm{iter}}=12$. For the sake of fairness, the outer loop iterations of the Turbo receiver\cite{Meng2017Low} are $N_{{\rm{iter}}}^{{\rm{out}}}=4$, and the inner loop iterations of EPA and LDPC are $N_{{\rm{iter}}}^{{\rm{EPA}}}=N_{{\rm{iter}}}^{{\rm{LDPC}}}=3$, as shown in the following Table \uppercase\expandafter{\romannumeral1}.

\begin{table}[htbp]
\centering
\renewcommand\arraystretch{1.2}
\caption{Scheduling of Turbo receiver and JDD receiver.}
\setlength{\tabcolsep}{1.2mm}
\begin{tabular}{|c|c|c|c|c|c|c|c|c|c|c|c|c|c|c|}
\hline
\multicolumn{2}{|c|}{Iterations}
& 1 & 2 & 3 & 4 & 5 & 6 & ... & 19 & 20 & 21 & 22 & 23 & 24 \\ \hline
\multirow{2}{*}{\begin{tabular}[c]{@{}c@{}}Turbo\\ receiver\end{tabular}}
 & EPA  &\checkmark &\checkmark &\checkmark &--  &--  &--  & ... &\checkmark &\checkmark &\checkmark &--  &--  &-- \\ \cline{2-15}
 & LDPC &--  &--  &--  &\checkmark &\checkmark &\checkmark &... &-- &-- &--  &\checkmark  &\checkmark  &\checkmark  \\ \hline
\multirow{2}{*}{\begin{tabular}[c]{@{}c@{}}JDD\\ receiver\end{tabular}}
 & EPA  &\checkmark  &--  &\checkmark  &--  &\checkmark  &--  &... &\checkmark  &--  &\checkmark  &--   &\checkmark   &-- \\ \cline{2-15}
 & LDPC &--  &\checkmark  &--  &\checkmark  &--  &\checkmark  &...  &--   &\checkmark  &--  &\checkmark  &--   &\checkmark  \\ \hline
\end{tabular}
\end{table}

The complexities of the MPA detector and EPA detector in each iteration can be expressed as $\mathcal{O}\left( {I{N_r}N{M^{{d_f}}}} \right)$ and $\mathcal{O}\left( {I{N_r}NM{d_f}} \right)$\cite{Meng2017Low}, and the complexity of LDPC decoder is $\mathcal{O}\left( {K(2{d_v}J + (2{d_c}+1)(J - L))} \right)$\cite{R2018NOMA} using the normalized min-sum algorithm, where ${d_v}$ and ${d_c}$ denote the average column weight and the average row weight of the parity check matrix, respectively. Consequently, the total complexities of the Turbo receiver and the JDD receiver with EPA detector are $N_{{\rm{iter}}}^{{\rm{out}}} \cdot \left( {N_{{\rm{iter}}}^{{\rm{EPA}}} \cdot {\cal O}\left( {{\rm{EPA}}} \right) + N_{{\rm{iter}}}^{{\rm{LDPC}}} \cdot {\cal O}\left( {{\rm{LDPC}}} \right)} \right)$ and ${N_{{\rm{iter}}}} \cdot \left( {\mathcal{O}\left( {{\mathrm{EPA}}} \right) + \mathcal{O}\left( {{\mathrm{LDPC}}} \right)} \right)$, respectively.

\begin{figure}[htbp]
  \centering
  \includegraphics[scale=0.5]{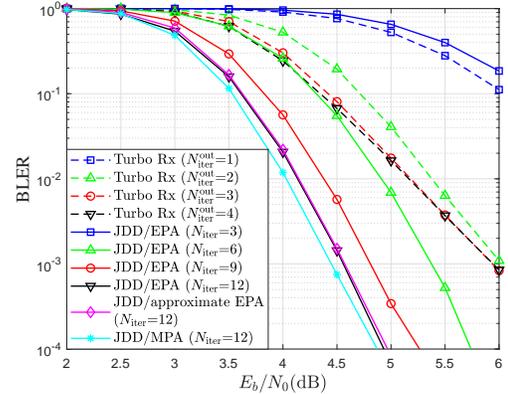}
  \caption{BLER performance comparisons between the JDD receiver and the Turbo receiver.}
\end{figure}

Fig. 2 presents the block error rate (BLER) performance comparisons between the different receiver schemes.
Here the threshold value is set as $\beta=1$ for the simulation. The JDD/EPA ($N_{{\rm{iter}}}=6$) can achieve the performance close to the Turbo receiver ($N_{{\rm{iter}}}^{{\rm{out}}}=4$) with only 50\% the computational complexity and improved error rate in the high SNR regime. On the other hand, the JDD/EPA ($N_{{\rm{iter}}}=12$) enjoys about 1.4 dB performance gain at BLER = ${10^{ - 3}}$ compared to that of the Turbo receiver ($N_{{\rm{iter}}}^{{\rm{out}}}=4$), when both receivers have more or less the same computational complexity. In addition, the BLER performance of the JDD/approximate EPA receiver well aligns with that of the JDD/EPA receiver with the same iterations.
And the JDD/EPA receiver only suffers from about 0.1 dB performance loss at BLER = ${10^{ - 3}}$ compared to the JDD/MPA receiver \cite{Han2016High}. In short, the proposed EPA based JDD receiver achieves improved BLER performance and faster convergence rate compared to the Turbo receiver.

\begin{figure}[htbp]
  \centering
  \includegraphics[scale=0.5]{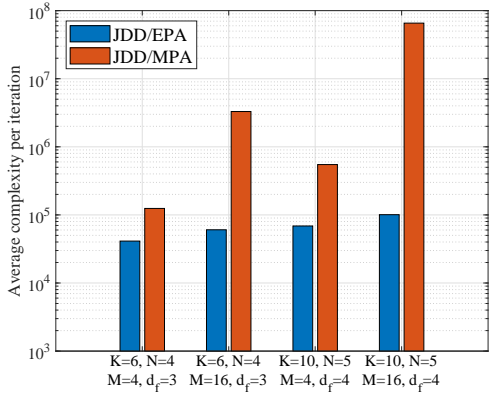}
  \caption{The average complexity per iteration comparisons between the JDD/EPA and JDD/MPA receivers.}
\end{figure}
Fig. 3 depicts the comparisons of the average complexity per outer iteration between the JDD/EPA and JDD/MPA receivers with different SCMA parameters, which can be expressed as ${I{N_r}NM{d_f}}+{K(2{d_v}J + (2{d_c}+1)(J - L))}$ and ${I{N_r}N{M^{{d_f}}}}+{K(2{d_v}J + (2{d_c}+1)(J - L))}$, respectively. It is clear that the complexity of the JDD/EPA is significantly lower than that of the JDD/MPA especially when the SCMA codebook size $M$ and/or the degree of active users superposition $d_f$ on a given resource element are larger. That is because the EPA detection has linear complexity with $M$, $d_f$ and $N_r$, whilst the complexity of the MPA detection is exponential with $M$ and $d_f$.

\begin{figure}[htbp]
  \centering
  \includegraphics[scale=0.5]{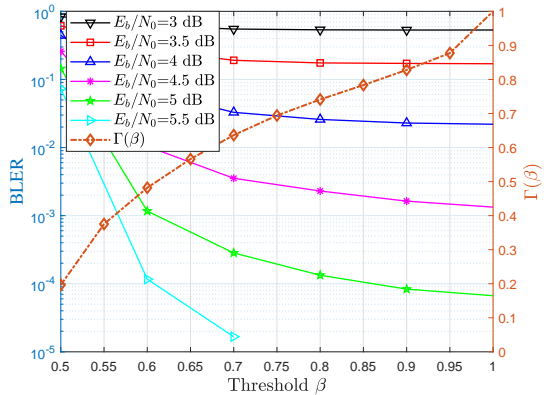}
  \caption{BLER performance and $\Gamma(\beta)$ of the JDD/EPA receiver with the different threshold values.}
\end{figure}

Fig. 4 shows that BLER performance becomes deteriorated with the decreasing threshold value $\beta$ under different $E_b/N_0$. The complexity reduction rate $\Gamma(\beta)$ is also presented, which is defined as the ratio of the effectively updated SCMA blocks and the total calculated SCMA blocks, in order to measure the reduced EPA detection complexity. For example, when the threshold value $\beta=0.75$, the EPA detection complexity of all SCMA blocks can be reduced by about 30\% with slight performance loss, hence helping strike a good balance between the error rate performance and complexity.

\section{Conclusions}
In this paper, we have presented an efficient iterative receiver for uplink LDPC coded SCMA system by combining an improved EPA detector with JDD receiver. Compared to the conventional Turbo receiver, the proposed JDD receiver is capable of achieving significant performance gains and accelerating the convergence rate. Our numerical results have shown that the proposed JDD/EPA receiver yields error rate performance close to that of the JDD/MPA receiver. Furthermore, by setting an appropriate threshold value in EPA detection, a significant amount of message passing operations can be saved, thus leading to a lower detection complexity without noticeable performance degradation. This demonstrates its potential advantage in practical implementations.


\end{document}